\documentclass[aps,pra,preprint,groupedaddress,showkeys,floatfix]{revtex4}

\usepackage[dvips]{graphicx}

\begin{document}
\title{Free-free transitions in a bichromatic field of frequencies $\omega $
and $2\omega$ at moderate field intensities}

\author{Aurelia Cionga and Gabriela Zloh}
\affiliation{Institute for Space Sciences, P.O. Box MG-36, Bucharest-M\u agurele,Bucharest, R-76900 Romania}

\begin{abstract}
Free-free transitions in laser-assisted electron-hydrogen scattering
in a bichromatic field of frequencies $\omega $ and $2\omega $
are studied at moderate intensities for fast projectiles.
A hybrid approach is used, in which the field-projectile interaction
is described exactly but the field-target one is described by second order
perturbation theory; the projectile-target interaction is treated in the
first Born approximation.
The adopted description of the target enables a consistent study of the leading process to each of the considered sidebands.
Numerical results are presented for the angular distributions in a geometry
and at frequencies for which the target dressing is important.
The influence of the relative phase between the fields is investigated, too.
\end{abstract}

\maketitle

\section{Introduction}
In the last years it has been observed that laser-assisted and laser-induced
processes may be considerably modify when they take place in a bichromatic
field. In this context, a special attention is shown to the case
of commensurate frequencies, in connection with high harmonic
generation experiments.
Theoretical investigations on free-free transitions
in laser-assisted electron-atom scattering in a bichromatic field have recently
been published. We quote here the paper by Varr{\'o} and Ehlotzky \cite{var},
where the development of the topic is presented.
The early results were obtained
for low frequencies, neglecting the dressing of the target \cite{ve1}.
They represent generalizations of the Bunkin and Fedorov formula \cite{fed},
but there are results \cite{cim} which go beyond the first Born approximation
in the scattering potential, extending the Kroll-Watson formula \cite{k-w} to
a bichromatic field.
However, perturbative calculations for both monochromatic
\cite{maq}-\cite{br1} and bichromatic fields \cite{pra}
have shown that the dressing of the target by the radiation field
plays an important role when the field frequency is no longer small.
Varr{\'o} and Ehlotzky \cite{var} were the first to take into account the
effect of target dressing in free-free transitions
in a bichromatic field at moderate field intensities for
fast projectiles.
They extend the approach introduced by Byron and Joachain \cite{joa} to deal
with the same problem in the monochromatic case.
In this treatment the interaction between the projectile and the field is the only
one to be treated exactly; the other two, namely the interaction between
the field and the bound electron and the projectile-target interaction are
treated in the framework of perturbation theory.
In Ref.\cite{var}
the laser-atom interaction is described by first order perturbation theory.

It is the aim of this paper to investigate free-free transitions
in a bichromatic field that is the superposition of the fundamental and of the
first harmonic at {\it moderate} field intensities for {\it fast} projectile.
We are interested by this process in the case of atomic hydrogen in the
ground state for frequencies which are
large enough to produce important dressing effects.
Section 2 is devoted to the approach adopted in this work;
the formalism used by Varr{\'o} and Ehlotzky \cite{var}
is extended: the description of the target dressing is improved
by including second order corrections in the electromagnetic field.
In Section 3 we discuss in detail the domain of small scattering angles
showing that, in the limit of small momentum transfer and
as long as the field intensities remain moderate,
the nonperturbative approach that we use here reduces to
a perturbative one.
In the case of the considered bichromatic field we claim that our calculations,
which include second-order corrections to the atomic state,
allow us to describe for each of the first four pairs of sidebands
at least the leading process consistently
(taking into account all the involved Feynman diagrams).
Section 4 contains the numerical results obtained for fast projectile,
$E_i$=100 eV, and two values of the fundamental frequency, namely
$\omega =$ 1.17 eV and $\omega =$ 4 eV. Angular distributions and
phase effects are discussed for different intensities in the moderate regime
pointing out the role played by the second order dressing of the target.

\section{Basic Formula}
This work is based on the assumption that at moderate field intensities
(significantly lower than the atomic unit),
the field-atom interaction can be described
using time-dependent perturbation theory \cite{joa}.
We use {\it second order} perturbation theory to
describe the hydrogen ground state in the presence of the bichromatic
field
\begin{equation}
\vec{\cal A}\left( t \right) = \vec A_1 \cos \omega_1t +
			\vec A_2 \cos \left( 2\omega_1 t + \varphi \right),
\label{har}
\end{equation}
which is the superposition of the fundamental, of frequency
$\omega_1 $, and of the first harmonic, of frequency
$\omega_2 = 2 \omega_1$.
$\;\vec A_k =\vec \varepsilon_k \sqrt{I_k}/\omega_k$
is the vector potential of the component $k$, with $k=1,2$.
$\;\vec \varepsilon_k $ denotes the polarization vector
and $I_k$ the intensity of that component, $\varphi $ is
the phase difference between the two components.

According to Florescu {\it et al} \cite{fhm}, one can write an approximate
solution for a Coulomb electron in an electromagnetic field as follows
\begin{equation}
|\Psi_{1} \left( t \right) > = e^{-i{\rm E }_{1s}t}
	\left[|\psi_{1s}> + |\psi_{1s}^{(1)}>+|\psi_{1s}^{(2)}> \right] ,
\label{fun}
\end{equation}
where $|\psi_{1s}>$ is the unperturbed ground state of hydrogen,
of energy ${\rm E}_{1s}$, and $|\psi_{1s}^{(1,2)}>$ denote first
and second order corrections, respectively. In agreement with
Refs.\cite{fhm} and \cite{f-m}
these corrections can be written in terms of the linear response
\begin{equation}
\vert{\vec w}_{1s}(\Omega)>=- G_C(\Omega)\vec P  \vert \psi_{1s}>,
\end{equation}
and of the second order tensor
\begin{equation}
| w_{ij,1s} (\Omega^{\prime}, \Omega ) > =
               G_C ( \Omega^{\prime} ) P_i G_C( \Omega ) P_j |\psi_{1s} >.
\end{equation}
Here $G_C ({\Omega})$ is the Coulomb Green's function
and $\vec P$ is the momentum operator of the bound electron.
For the bichromatic field (\ref{har}), there are twelve values of the parameter
of the Green functions which are necessary in order to write the approximate
solution (\ref{fun}), namely
\begin{eqnarray}
\Omega^{\pm}= {\rm E}_{1s} \pm (\omega_{1} - \omega_{2}), \label{omdif}
		&\quad &
	\Omega^{'\pm}= {\rm E}_{1s} \pm (\omega_{1}+\omega_{2}), \\
\Omega_{k}^{\pm}= {\rm E}_{1s} \pm \omega_{k},  &\quad &
	\Omega^{'\pm}_{k}= {\rm E}_{1s} \pm 2\omega_{k} .
\label{omeq}
\end{eqnarray}

On the other hand, the interaction between the bichromatic field and the
projectile is treated exactly by using the Volkov-type solution
\begin{eqnarray}
\chi_{\vec p}(\vec r,t) & = &\frac{1}{(2\pi)^3}
	 \exp \left\{-iE_pt+i\vec p \cdot \vec r
\right. \nonumber \\
&& \hspace*{1.5cm} \left.
	    - i\vec p \cdot [{\vec \alpha_1}(t)+ \vec \alpha_2(t)]\right\}.
\label{fe}
\end{eqnarray}
where
\begin{eqnarray}
{\vec \alpha_1 } (t)&=&\vec{\varepsilon_1}\alpha_{01}\sin (\omega_1 t),
		\nonumber\\
{\vec \alpha_2 } (t)&=&\vec{\varepsilon_2}\alpha_{02}\sin (\omega_2 t+\varphi) .
\label{alpha}
\end{eqnarray}
$\vec r$ is the position, $\vec p$ the momentum, and
$E_p$ the energy of the free electron;
$\alpha_{0k}=\sqrt{I_k}/\omega_k^2$ denotes the amplitude of the quiver motion
for the component $k$ of the field (\ref{har}).

We restrict ourselves to the domain of high scattering energies, where first
Born approximation in the scattering potential
may be reliable. Neglecting the exchange effects,
we describe this interaction by the static potential, $V(r, R)$,
and the scattering matrix element is given by
\begin{equation}
S^{B1}_{if}=-i\int_{-\infty}^{+\infty}
		dt<{\chi}_{{\vec p}_f} (t)\Psi_{1}(t)
                                 |V|
                {\chi}_{{\vec p}_i} (t)\Psi_{1}(t) >,
\label{GZ}
\end{equation}
where $\Psi_{1}$ and ${\chi}_{{\vec p}_{i,f}}$ are written using
Eqs.(\ref{fun}) and (\ref{fe}).

In the presence of the radiation field (\ref{har}) the electron scattered on
hydrogen may gain or loose an energy equal to $n \omega_1$, such that
$E_f=E_i+n\omega_1\equiv E_i+n_1\omega_1 + n_2\omega_2 $,
where $E_{i(f)}$ is the initial (final) energy of the projectile and
$n_k$ is the net number of photons $\omega_k$ exchanged
(absorbed or emitted) by the colliding system and the $k$ component
of the field.
The energy spectrum of the scattered electron consists therefore
of the elastic line, corresponding to $n=0$, and of a number of sidebands,
each pair of sidebands corresponding to the same $|n|$.
The differential cross section for any process in which
the energy of the projectile is {\it modified} by $n\omega_1$
is written as
\begin{equation}
\frac{d{\sigma}(n)}{d\Omega} = {(2\pi)}^4 \frac{p_f{(n)}}{p_i}
					{| T_{if}(n) |}^2 ,
\label{sec}
\end{equation}
where the transition matrix element, related to the $S$-matrix (\ref{GZ}),
has the following general structure
\begin{equation}
T_{if}(n) =T_n^{(0)}+T_n^{(1)}+T_n^{(2)}.
\label{gen}
\end{equation}

The first term,
\begin{equation}
T_n^{(0)}=B_n(a_1,a_2,\varphi) <\psi_{1s}|F(\vec q)|\psi_{1s}> ,
\label{t0}
\end{equation}
might be seen as the equivalent of Bunkin-Fedorov formula for
a bichromatic field \cite{ve1}. In the previous equation
$F(\vec{q})$ is the form factor operator
\begin{equation}
F(\vec{q}) = \frac{1}{2 \pi^2 q^2}\left[ \exp{(i \vec{q} \cdot \vec {r} )} - 1
\right]
\label{ff}
\end{equation}
and $B_n(a_1,a_2,\varphi)$ are generalized Bessel functions
\begin{equation}
B_n(a_1,a_2,\varphi)=\sum_{n_2=-\infty}^{\infty}
			 J_{n-2n_2}(a_1)J_{n_2}(a_2)e^{-i n_2 \varphi};
\label{genb}
\end{equation}
${\vec q}(n)$ is the momentum transfer of the projectile,
$ \vec q (n)= \vec p_i- \vec p_f(n)$ such that $ p_f^2/2=p_i^2/2+n\omega_1$, and
the arguments of the two Bessel functions are given by
\begin{equation}
a_k(n) = \alpha_{0k} \; {\vec \varepsilon}_k \cdot  {\vec q}(n).
\end{equation}
If the dressing of the target is neglected then $T_{if}(n) = T_n^{(0)}$ and
the generalized Bessel function, $B_n(a_1, a_2, \varphi)$, contains all the
field dependence of the transition matrix in the same way in which
$J_{n_1}(a_1)$ would do it in Bunkin-Fedorov formula \cite{fed}
for a monochromatic field of frequency $\omega_1$.

The other two terms in Eq.(\ref{gen}) are due to the modification
of the atomic state in the bichromatic field.
The second term, $T_n^{(1)}$, is connected to the first order
corrections to the atomic state:
one of the $N$ photons exchanged between
the field (\ref{har}) and the colliding system  interacts
with the bound electron. We note that $N=|n_1|+|n_2| \neq n=n_1+ 2 \; n_2$.
This photon may have the energy $\omega_1$ or $\omega_2$,
it may be emitted or absorbed, therefore the general
structure of $T_n^{(1)}$ is given by
\begin{eqnarray}
T_n^{(1)}&=& - \sum_{k=1}^2 \frac{\alpha_{0k} \omega_k}{2}
	\left[ B_{n+k}\;
		f_k^+ \left( \varphi \right) \;
			{\cal M}_{at}^{(I)} \left( \Omega_k^-\right)
	\right. \nonumber \\
	&& \hspace*{1.8cm}\left.  + B_{n-k} \;
		f_k^- \left( \varphi \right)\;
			{\cal M}_{at}^{(I)} \left( \Omega_k^+ \right) \right],
\label{t1}
\end{eqnarray}
where $f_k^{\pm}$ is a function of the relative phase between the fields:
\begin{equation}
f_k^{\pm}\left( \varphi \right)=\left\{ \begin{array}{ll}
	1 & \mbox{if $k$=1} \\
	\exp\left({\pm i \varphi}\right) & \mbox{if $k$=2}.
	\end{array}
	\right.
\end{equation}
${\cal M}_{at}^{(I)}$ denotes the following matrix elements involving
atomic states
\begin{eqnarray}
{\cal M}_{at}^{(I)}\left( \Omega_k^{\pm}\right) & = &
	<\psi_{1s}|F(\vec q)|
		{\vec \varepsilon_k \cdot \vec w}_{1s}(\Omega_{k}^\pm)>
		\nonumber \\ & + &
	<{\vec \varepsilon_k \cdot \vec w}_{1s}(\Omega_k^{\mp})
		|F(\vec q)|\psi_{1s}>,
\label{defm1}
\end{eqnarray}
its significance will became clear in the next section.

Varr{\'o} and Ehlotzky \cite{var} studied free-free transitions in a
bichromatic field taking into account only the first order corrections
to the atomic ground state, the corresponding
transition matrix element being given by the sum of
$T_n^{(0)}$ and $T_n^{(1)}$.
Moreover, in Ref.\cite{var} the atomic matrix elements (\ref{defm1}) were
evaluated in the closure approximation. Very recently, some results were
published \cite{mil}, which are based on the Sturmian representation of the
Coulomb Green's function.

Including second order corrections in the approximate description
of the atomic ground state (\ref{fun}) we have to evaluate a third term,
$T_n^{(2)}$, in Eq.(\ref{gen}). In this term two of the
$N$ photons exchanged between the fields and the colliding system
interact with the bound electron:
\begin{eqnarray}
&&T_n^{(2)} = \sum_{k=1}^2 \frac{ \alpha_{0k}^2 \omega_k ^2 }{4}
	\left\{ B_{n+2k}
		\left[f_k^+ \left( \varphi \right)\right]^2
		{\cal M}_{at}^{(II)}
			\left( \Omega_k^{\prime -},\Omega_k^-\right)
	\right. \nonumber \\
	&& \hspace*{2cm}
	+ B_{n-2k}
		\left[f_k^- \left( \varphi \right)\right]^2
		{\cal M}_{at}^{(II)}
			\left( \Omega_k^{\prime +},\Omega_k^+ \right)
			  \nonumber \\
	&& \hspace*{2cm}
	\left. + B_{n}
	            \left[\widetilde {\cal M}_{at}^{(II)}
	                  \left( {\rm E}_{1s},\Omega_k^- \right)
	           +\widetilde {\cal M}_{at}^{(II)}
	                  \left( {\rm E}_{1s},\Omega_k^+ \right)\right]
		\right\} \nonumber \\
&&+ \frac{\alpha_{01} \alpha_{02} \omega_1 \omega_2 }{4}
	\left[ B_{n+3}
		\; f_2^+ \left( \varphi \right)\;
		 {\cal N}_{at}^{(II)}
			\left( \Omega^{\prime -},\Omega_1^-,\Omega_2^-\right)
	\right. \nonumber \\
	&& \hspace*{2cm} + B_{n-3}
		\;f_2^- \left( \varphi \right) \;
		{\cal N}_{at}^{(II)}
		\left( \Omega^{\prime +}, \Omega_1^+, \Omega_2^+ \right)
\nonumber \\
&& \hspace*{2cm} +
 		B_{n+1}
			\; f_2^+ \left( \varphi \right)\;
			{\cal N}_{at}^{(II)}
				\left( \Omega^+, \Omega_1^+, \Omega_2^- \right)
	\nonumber \\
&& \hspace*{2cm}\left. +
		B_{n-1}
			\; f_2^- \left( \varphi \right) \;
		 	{\cal N}_{at}^{(II)}
				\left( \Omega^-, \Omega_1^-, \Omega_2^+\right)
		\right].
\label{t2}
\end{eqnarray}
Two types of atomic matrix elements appear in Eq.(\ref{t2});
they are related to the exchange of {\it identical} photons
\begin{eqnarray}
&&{\cal M}_{at}^{(II)}\left( \Omega_k^{\prime \pm}, \Omega_k^{\pm} \right) =
\nonumber \\ && \hspace*{0.25cm}
	\sum_{j,l=1}^{3} \varepsilon_{kj} \varepsilon_{kl}
\left[ < \psi_{1s}|F(\vec q)|
	w_{lj,1s} \left( \Omega_k^{\prime \pm}, \Omega_{k}^{\pm}\right) >
	\right. \nonumber \\
&& \hspace*{1.5cm}+<w_{j,1s}(\Omega_k^{\mp})|F(\vec q)|
	{w}_{l,1s}(\Omega_k^{\pm})>
\nonumber \\
&& \hspace*{1.5cm} +\left.
	< w_{lj, 1s} \left( \Omega_k^{\prime \mp}, \Omega_k^{\mp} \right)
	|F(\vec q)|\psi_{1s}> \right],
\label{defm}
\end{eqnarray}
or of {\it different} photons
\begin{eqnarray}
&&{\cal N}_{at}^{(II)}
	\left( {\tilde \Omega}^{\pm}, \Omega_m^{\pm}, \Omega_n^{\pm} \right)
=\left( 1+{\cal P}_{mn} \right)
\nonumber \\ && \hspace*{0.1cm} \times
	\sum_{j,l=1}^{3} \varepsilon_{mj} \varepsilon_{nl}
\left[ < \psi_{1s}|F(\vec q)|
	w_{lj,1s} \left( {\tilde \Omega}^{\pm}, \Omega_{m}^{\pm}\right) >
	\right. \nonumber \\
&&  \hspace*{1.8cm} + <w_{j,1s}(\Omega_m^{\mp})|F(\vec q)|
	{w}_{l,1s}(\Omega_n^{\pm})>
\nonumber \\ &&\hspace*{1.8cm} + \left.
< w_{lj, 1s} \left( {\tilde \Omega}^{\mp}, \Omega_m^{\mp} \right)
	|F(\vec q)|\psi_{1s}> \right],
\label{defn}
\end{eqnarray}
where $k, m$, and $n$ take the values $1$ and $2$, but $m \neq n$.
$\tilde \Omega^{\pm}$ is a generic
notation for $\Omega^{\pm}$ or $\Omega^{\prime \;\pm}$ defined in
Eq.(\ref{omdif}).
${\cal P}_{mn}$ denotes a permutation operator that interchanges the subscripts
$m$ and $n$.
$\widetilde {\cal M}_{at}^{(II)}$ is built from Eq.(\ref{defm}) using the tensor
$\widetilde {w}_{ij,1s}$,
defined in Ref.\cite{fhm}, instead of
$w_{ij,1s}$.
For the sake of simplicity, the arguments of the generalized Bessel functions
$B_n \left( a_1, a_2, \varphi \right) $ are omitted in Eqs.(\ref{t1},\ref{t2})
as well as the $q$-dependence of the atomic matrix elements in
Eqs.(\ref{defm1}, \ref{defm}-\ref{defn}).
Different methods may be used to evaluate the atomic matrix elements in
Eqs.(\ref{defm1}, \ref{defm}-\ref{defn}), which are already known from
perturbative calculations involving one \cite{maq},\cite{ac1}
or two photons \cite{br1},\cite{pra}.
We use in the numerical evaluations reported here the analytic
expressions of these atomic matrix elements
as series of hypergeometric functions \cite{pra},\cite{ac1}.

\section{Small scattering angles}

For small values of the arguments $a_1$ and $a_2$, the Bessel functions
$J_{n_1}\left(a_1\right)$ and $J_{n_2}\left(a_2\right)$
have approximate values given by
\begin{equation}
J_n(x) \simeq \frac{1}{n!}\left(\frac{x}{2} \right)^n
\label{lim}
\end{equation}
and the generalized Bessel functions, $B_n(a_1,a_2,\varphi)$,
can be approximated by a sum of a few terms, as it is shown below.
In general, the arguments $a_1$ and $a_2$ are small for low field intensities
where one recovers perturbative results.
It is important to note that,
in those geometries for which the polarization vector is almost orthogonal
to the momentum transfer of the projectile,
these arguments are small for every intensity since
$ a_{1,2} \sim \vec \varepsilon_{1,2} \cdot \vec q $.

We focus here the attention on the case of linear, identical polarizations
${\vec \varepsilon}_1 = {\vec \varepsilon}_2 \equiv \vec \varepsilon$,
for the geometry in which the initial momentum 
is parallel to the polarization vector, defining the $Oz$-axis.
Based on the previous remark, we show that
in this geometry, for small scattering angles such that
$\vec \varepsilon \cdot \vec q \ll 1$ and
as long as the intensities remain moderate,
perturbation theory might represent a sensible
treatment of free-free transitions.
For the bichromatic field (\ref{har}) the approach developed here allows us
to study the leading processes to the first four pairs of sidebands,
$|n|\leq 4$ in Eq.(\ref{sec}).
In addition, for the first two pairs of sidebands, the first high order
processes are also taken into account.
In all these cases, all the involved Feynman diagrams are included,
enabling us to account consistently for the effects of the target dressing.
We remind that in this geometry  the dressing of the target,
connected to the last two terms in Eq.(\ref{gen}), is important
at small scattering angles.
For large scattering angles $a_{1,2}$ might be large and
perturbation theory may no longer be valid if the fields are high enough.
The dressing of the target is negligible for large scattering angles
therefore the dominant contribution comes from the electronic diagrams,
related to the term $T_n^{(0)}$;
no significant differences between results based on Eq.(\ref{gen}) and
Ref.\cite{ve1} are to be expected in this domain.

\subsection{First pair of sidebands: $n=\pm1$}
To prove that, in the geometry chosen here, perturbation theory is
a reliable treatment at small scattering angles let us see what is
the behavior of the transition matrix element (\ref{gen}) in the
limit $a_{1,2} \ll 1$.
The key point in this analysis is the behavior of the
generalized Bessel functions (\ref{genb}) when use is made of Eq.(\ref{lim}).
Our goal
is to describe consistently at least the leading contribution to each sideband
with $|n| \leq 4$, therefore we keep systematically second order terms in the
fields. To fulfill the same goal for higher $|n|$ it would be necessary to
add higher order corrections to the atomic state (\ref{fun}).

For $n=1$ we discuss in some detail each of the three terms
in Eq.(\ref{gen}). Keeping only second order contributions in the fields,
one has
\begin{eqnarray}
B_1 \left( a_1, a_2, \varphi \right) & \simeq &
	J_1(a_1)J_0(a_2) + J_{-1}(a_1) J_1(a_2) e^{-i\varphi}
	\nonumber \\
& \simeq & \frac{a_1}{2} - \frac{a_1 a_2}{4} e^{-i\varphi },
\label{b1}
\end{eqnarray}
which leads to the following form of the electronic term
\begin{equation}
T_1^{(0)} \simeq \left( \frac{a_1}{2} - \frac{a_1 a_2}{4} e^{-i\varphi} \right)
	<\psi_{1s}|F(\vec q)|\psi_{1s}>.
\label{t00}
\end{equation}
In this equation we neglect all the terms of order three or higher, namely
$a_1^j a_2^{s-j}$ with $s\geq 3$ and $j=\overline{0,s}$.
Each of the two terms in Eq.(\ref{t00}) is connected to a specific
quantum path. In the first one
the projectile absorbs one photon of energy $\omega_1$. It is described by two
Feynman diagrams shown in Fig.1(a).
In the second term
the projectile absorbs a photon of energy $\omega_2 = 2 \omega_1$
and emits a photon of energy $\omega_1$, the scattered electron has the same
final energy as in the previous case:
$ E_f=E_i+\omega_1$.
This term is of the second order in the fields and
it is described by six Feynman diagrams. Only three of them are shown
in Fig.1(b),
the other three are obtained by interchanging $\omega_1$ and $\omega_2$ in time
(interchanging their order on the vertical lines).

\begin{figure}
\includegraphics[width=3in,angle=0]{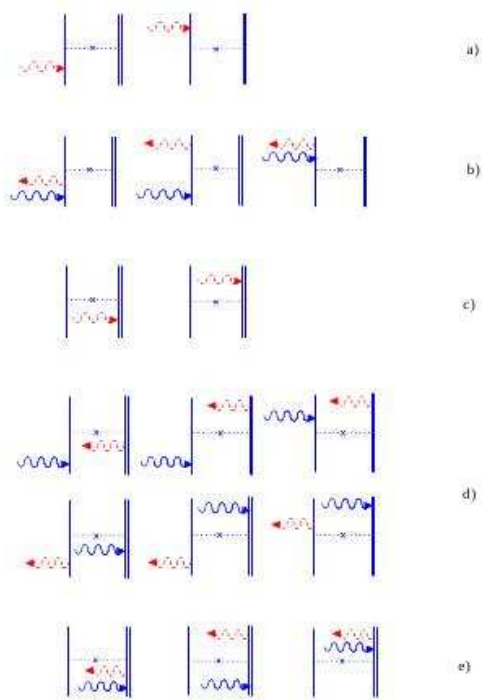}
\caption{(a) Feynman diagrams corresponding to
one photon absorption by the projectile. The single line represents the
free- and the double one the bound-electron. The horizontal line
denotes the projectile-target interaction.
 (b) Feynman diagrams describing the absorption of $\omega_2$
and the emission of $\omega_1$ by the free electron.
 (c) Same as Fig.1(a) but for the bound electron.
(d) Feynman diagrams in which each electron
interacts with one photon.
(e) Same as Fig.1(b) but for the bound electron.}
\end{figure}

A similar technique is used to find out the behavior of the second term
in Eq.(\ref{gen}).
In the limit we are interested in,
only three generalized Bessel functions have a contribution to Eq.(\ref{t1}),
namely  $B_0$, $B_{-1}$, and $B_2$.
The fourth one, $B_3$, has a leading term which is of
higher order in the fields, therefore it is neglected.
Finally, one gets
\begin{eqnarray}
&&T_1^{(1)}  \simeq
 - \frac{1}{2}
	\left\{ \alpha_{01} \omega_1
			{\cal M}_{at}^{(I)} \left( \Omega_1^+ \right)
\right. \label{t11} \\
&&+ \left. \alpha_{01}\alpha_{02}
		\frac{\vec \varepsilon \cdot \vec q}{2}  e^{-i \varphi }
	\left[ \omega_1{\cal M}_{at}^{(I)} \left( \Omega_1^- \right)
		- \omega_2{\cal M}_{at}^{(I)} \left( \Omega_2^+ \right)
		 	\right]\right\}.
\nonumber
\end{eqnarray}
The first term involves one photon $\omega_1$ that is absorbed by the
atomic electron. The corresponding two Feynman diagrams can be seen in Fig.1(c).
The second term involves two photons of different colors:
a photon $\omega_1$ is emitted by the bound/free electron and
an other photon, $\omega_2$, is absorbed by the free/bound electron.
In Fig.1(d) only
six Feynman diagrams are shown, the other six are obtained by interchanging
in time $\omega_1$ and $\omega_2$.

In the limit $a_{1,2} \ll 1$, the last term in Eq.(\ref{gen})
has  only one contribution, given by the last line in Eq.(\ref{t2})
\begin{equation}
T_1^{(2)}
\simeq  \frac{1}{4} \alpha_{01} \alpha_{02} \omega_1 \omega_2
	\; e^{-i \varphi } \;
		{\cal N}_{at}^{(II)}
			\left( \Omega^-, \Omega_1^-, \Omega_2^+\right).
\label{t22}
\end{equation}
It represents a second order term in which both photons,
$\omega_1$ and $\omega_2$, interact with the
atomic electron. Three of the six corresponding Feynman diagrams
are shown in Fig.1(e).

Adding together the approximate forms in Eqs.(\ref{t00}-\ref{t22}),
one gets finally the following form of the transition matrix element
\begin{eqnarray}
T(1)
&\simeq & \frac{\alpha_{01}}{2}
	\left[ \vec \varepsilon \cdot \vec q
		<\psi_{1s}|F(\vec q)|\psi_{1s}>
			-\omega_1{\cal M}_{at}^{(I)} \left( \Omega_1^+ \right)
				\right] \nonumber \\
&+& e^{-i \varphi } \; \frac{\alpha_{01}\alpha_{02}}{4}
	\left\{ -\left( \vec \varepsilon \cdot \vec q \right)^2
		<\psi_{1s}|F(\vec q)|\psi_{1s}> \right.\nonumber \\
&&
\hspace*{0.8cm} + \vec \varepsilon \cdot \vec q \; \left[
	\omega_2 {\cal M}_{at}^{(I)} \left( \Omega_2^+ \right)
	- \omega_1 {\cal M}_{at}^{(I)} \left( \Omega_1^- \right) \right]
	\nonumber \\ && \hspace*{0.8cm} + \left.
	\omega_1 \omega_2 {\cal N}_{at}^{(II)}
		\left( \Omega^-, \Omega_1^-, \Omega_2^+\right) \right\}
.
\label{app1}
\end{eqnarray}
The first {\bf line} in this equation represents the transition matrix element,
denoted by $T_a$,
describing one photon absorption (see Fig.2(a) and the corresponding
Feynman diagrams in Figs.1(a) and (c)).
The remaining part, $T_b$,
describes the two photon process shown in Fig. 2(b)
(see also Figs.1(b), (d) and (f)).

\begin{figure}
\includegraphics[width=3in,angle=0]{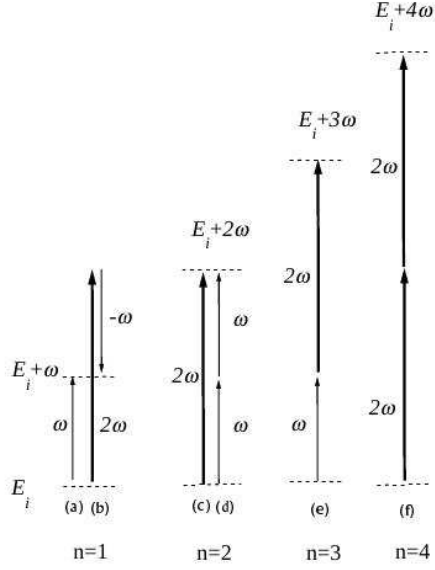}
\caption{{\bf (a-b)} Channels leading to the final energy
$E_f=E_i+n\omega_1$ with $n=1$.
{\bf (c-d)} Same as Fig.2(a) but $n=2$.
{\bf (e)} Same as Fig.2(a) but $n=3$.
{\bf (f)} Same as Fig.2(a) but $n=4$. }
\end{figure}

Accordingly, the differential cross section for the process in which
the scattered projectile has the energy
$E_f=E_i + \omega_1$
is approximated at small scattering angles by
\begin{eqnarray}
\frac {d\sigma (1)}{d\Omega} & \simeq & (2\pi)^4 \frac{p_f}{p_i}
	I_1 \left[ | {\cal T}_a |^2 +
		I_2 | {\cal T}_b |^2
\right. \nonumber \\ && \hspace*{2cm}+ \left.
			2\sqrt{I_2}\; Re \left({\cal T}_a^* {\cal T}_b
				e^{-i \varphi} \right) \right],
\label{sec1}
\end{eqnarray}
where we have chosen to display explicitly the dependence of the transition
matrix elements on the intensities and on the relative phase of the fields:
$T_a \equiv \sqrt{I_1} {\cal T}_a $ and
$T_b \equiv  \sqrt{I_1 I_2} {\cal T}_b \; e^{-i \varphi }$.

The same techniques may be used to study the case $n=-1$,
where the final energy is ${E_f} ={E_i}-\omega_1$ and the dominant process
is the stimulated emission of a photon $\omega_1$.

\subsection{Other sidebands: $n=\pm 2, \; \pm 3, \; \pm 4$}
The procedure presented before is now applied for the next sideband, $n=2$.
In this case the scattered electrons have the energy $E_f=E_i+2\omega_1$.
In the limit $a_{1,2} \ll 1$,
keeping again only second order terms in the fields,
the approximate transition matrix element
has a form similar to that given in Eq.(\ref{app1}), namely
\begin{eqnarray}
T(2)
&\simeq & e^{-i \varphi } \frac{\alpha_{02}}{2}
	\left[ \vec \varepsilon \cdot \vec q
		<\psi_{1s}|F(\vec q)|\psi_{1s}\!>
			\!-\omega_2{\cal M}_{at}^{(I)} \left( \Omega_2^+ \right)
				\right] \nonumber \\
& + &\frac{\alpha_{01}^2}{4}
	\left[ \frac{1}{2} \left( \vec \varepsilon \cdot \vec q \right)^2
		<\psi_{1s}|F(\vec q)|\psi_{1s}>
	\right. \nonumber \\ &-& \left.
	\vec \varepsilon \cdot \vec q \; \omega_1
		{\cal M}_{at}^{(I)} \left( \Omega_1^+ \right)
	+\omega_1^2{\cal M}_{at}^{(II)}
		\left( \Omega_1^{\prime +}, \Omega_1^+ \right) \right]
.
\label{app2}
\end{eqnarray}
The first {\bf line} represents the transition matrix element
describing the absorption of one photon $\omega_2$ in Fig.2(c)
and it is denoted by $T_c$.
The remaining part, denoted by $T_d$,
describes the absorption of two photons $\omega_1$, shown in Fig. 2(d).

The differential cross section for the process in which
the scattered projectile has the energy
$E_f=E_i + 2\omega_1$
is approximated at small scattering angles by
\begin{eqnarray}
\frac {d\sigma (2)}{d\Omega} & \simeq & (2\pi)^4 \frac{p_f}{p_i}
	I_2 \left[ | {\cal T}_c |^2 +
		\frac{I_2}{{\rm f}^2} | {\cal T}_d |^2
\right. \nonumber \\ && \hspace*{2cm}+ \left.
			2\frac{\sqrt{I_2}}{\rm f}
				Re \left({\cal T}_c^* {\cal T}_d
					e^{-i \varphi} \right) \right],
\label{sec2}
\end{eqnarray}
where we display again the explicit dependence on the intensities and phase:
$T_c \equiv \sqrt{I_2} {\cal T}_c $ and
$T_d \equiv I_1 {\cal T}_d \; e^{-i \varphi }$.
$\rm f$ denotes in Eq.(\ref{sec2}) the ratio between the intensities
of the harmonic and the fundamental,
$ {\rm f} =I_2/I_1 $.

When the limit $a_{1,2} \ll 1$ is  taken for $n=3$ and $n=4$ in Eq.(\ref{gen}),
one gets:
\begin{eqnarray}
T(3)& \simeq &
e^{-i \varphi } \; \frac{\alpha_{01}\alpha_{02}}{4}
	\left\{  \left( \vec \varepsilon \cdot \vec q \right)^2
			<\psi_{1s}|F(\vec q)|\psi_{1s}>
	\right. \nonumber \\
&&
	-\vec \varepsilon \cdot \vec q \left[
	\omega_1{\cal M}_{at}^{(I)} \left( \Omega_1^+ \right)
	+\omega_2{\cal M}_{at}^{(I)} \left( \Omega_2^+ \right) \right]
	\nonumber \\ && + \left.
	\omega_1\omega_2 {\cal N}_{at}^{(II)}
		\left( \Omega^{\prime +}, \Omega_1^+, \Omega_2^+\right)
			\right\}
\label{app3}
\end{eqnarray}
and
\begin{eqnarray}
T(4)
&\simeq & e^{-2i \varphi } \;\frac{\alpha_{02}^2}{4}
	\left[ \frac{1}{2} \left( \vec \varepsilon \cdot \vec q \right)^2
		<\psi_{1s}|F(\vec q)|\psi_{1s}>
	\right. \nonumber \\
	& - & \left. \vec \varepsilon \cdot \vec q \; \omega_2
		{\cal M}_{at}^{(I)} \left( \Omega_2^+ \right)
	+\omega_2^2 {\cal M}_{at}^{(II)}
		\left( \Omega_2^{\prime +}, \Omega_2^+ \right) \right]
.
\label{app4}
\end{eqnarray}
At small scattering angles, in the framework of the approach used here and
as long as we restrict ourselves to the second order in the electric fields,
the dominant process for $n=3$ is the absorption of two photons
of different colors (see Fig.2(e)).
For $n=4$, the dominant process is the absorption of two harmonic photons
(see Fig.2(f)).
In this approximation the
differential cross sections have simpler
dependences on the intensity of the fields:
\begin{eqnarray}
\frac {d\sigma (3)}{d\Omega} &\simeq &
	(2\pi)^4 \frac{p_f}{p_i} I_1 I_2 | {\cal T}_e |^2
,\label{sec3} \\
 \frac {d\sigma (4)}{d\Omega} & \simeq &
	(2\pi)^4 \frac{p_f}{p_i} I_2^2 | {\cal T}_f |^2.
\label{sec4}
\end{eqnarray}
The relative phase between the fields is not a relevant parameter if only the
leading process is taken into account.
\section{Numerical results and discussion}
We focus our attention on the study of free-free transitions
in electron-hydrogen scattering in the presence of a bichromatic field
for a high initial energy of the projectile, $E_i=$ 100 eV.
Its initial momentum, $\vec p_i$, is parallel to the polarization vectors
of the fields,
$\vec \varepsilon $, and defines the $Oz$-axis.
We are interested in the differential cross sections for processes in which
the energy of the scattered projectile is ${ E_f}={E_i} + n \omega_1$
with $n$ an integer such that $-4\leq n \leq4,\; n\neq 0$.
Having in mind the analysis presented in the previous section,
we investigate in detail the domain of small scattering angles,
where the dressing of the target is important.
The effect of the intensities of the two
components of the bichromatic field and that of their relative phase
is investigated
for two cases: $\omega $ = 1.17 eV and $\omega $= 4 eV.

\subsection{$\omega $= 1.17 eV}

Figs.3(a-d) show the differential cross sections calculated
for a bichromatic field that is the superposition of the fundamental
and the first harmonic of Nd:YAG laser ($\omega_1$=1.17 eV).
We treat here the case of equal intensities, $I_1=I_2=10^{12}$ W/cm$^2$,
and we consider that the fields are in phase, $\varphi=0$.
Full lines are used to represent the differential cross sections calculated
when all three terms are included in Eq.(\ref{gen}) and
Eqs.({\ref{t0}), (\ref{t1}), and (\ref{t2}) are used to compute them.
For $n=\pm 1; \pm 2$ the dominant process is of the first order in the field,
the frequency of the absorbed/emitted photon is $\omega_1$ when $n=\pm 1$ and
$\omega_2$ when $n=\pm2$. At this intensity,
the interferences due to second order processes (Figs.2(b) and (d))
are negligible; they do not play any significant role below
$I_{1,2}=10^{13}$ W/cm$^2$.
For $|n|\geq 3$ the dominant process involves two photons
(see Fig.2(e) and (f)).
In general, no significant deviations from the perturbative regime
exist in Figs.3 when $\theta <20^0$. The case $n=-4$ is an exception:
the minimum at $\theta=16^0$ is due to the interference between the terms in
Eq.(\ref{gen}), it is located close to a zero of  $B_{-4}(a_1, a_2, 0)$.
\begin{figure}
\includegraphics[width=5.5in,angle=0]{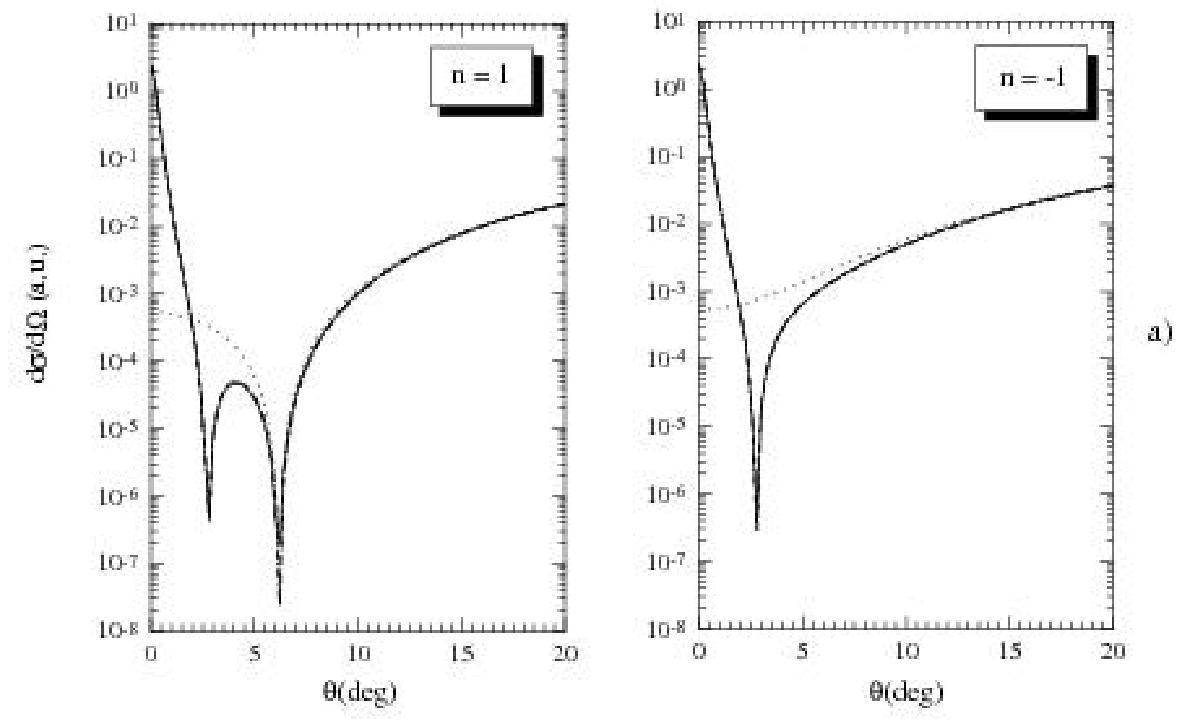}\\
\includegraphics[width=6in,angle=0]{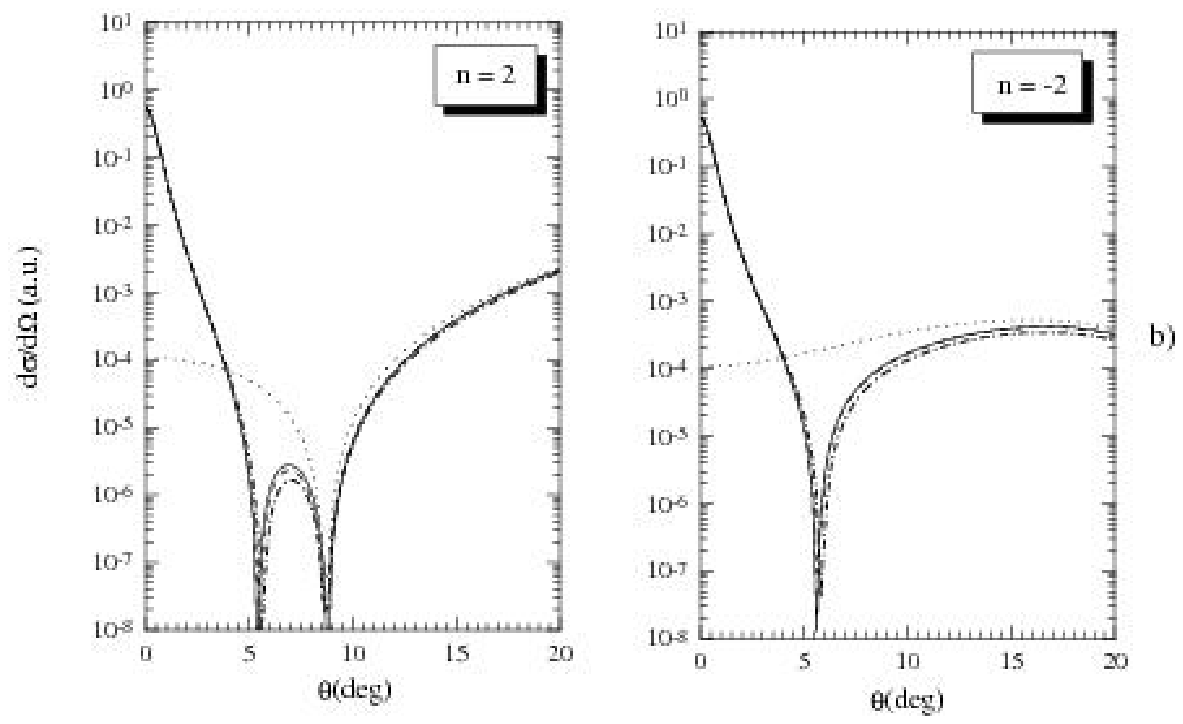}\\
\includegraphics[width=6in,angle=0]{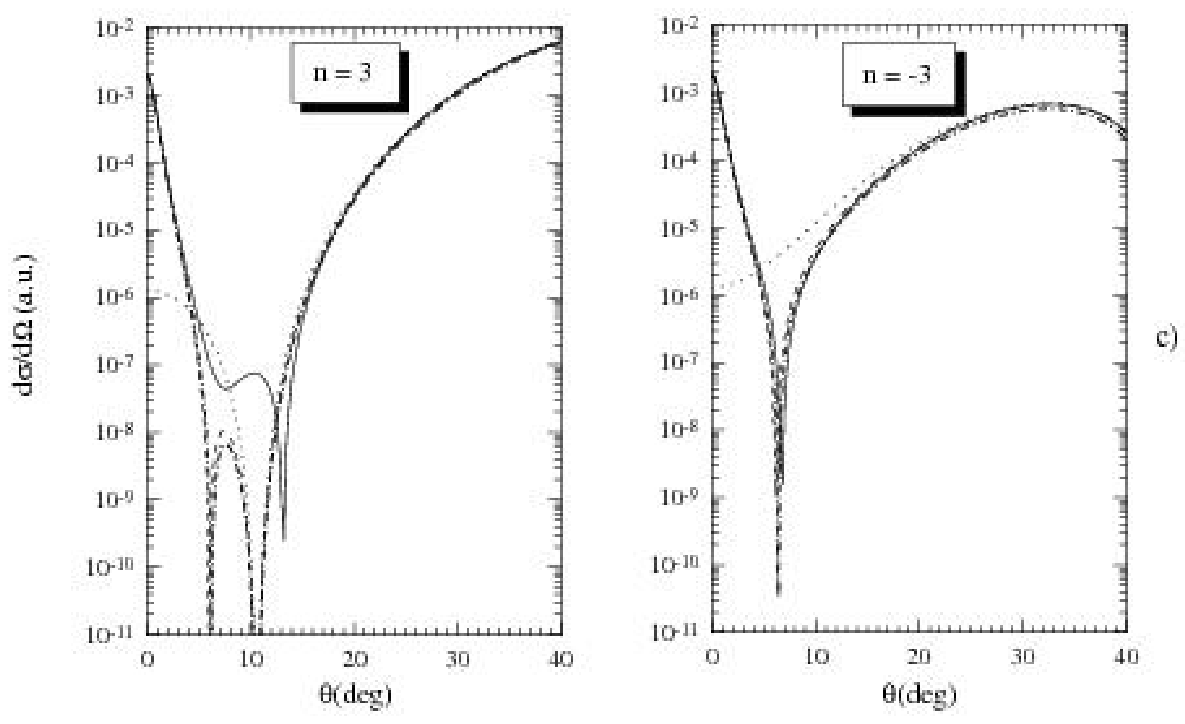}
\includegraphics[width=6in,angle=0]{nd3.eps}
\caption{{\bf (a)} Differential cross sections for $|n|=1$ as a function of the scattering
angle $\theta $ at the initial energy
$E_i=100$ eV for Nd:YAG laser ($\omega_1$=1.17 eV). $I_1= I_2=10^{12}$ W/cm$^2$
and the fields are in phase, $\varphi=0$. Full lines represent the results based on
Eq.(\ref{gen}), dotted lines include only first order dressing, dotted-dashed lines
correspond to the closure approximation, dotted lines do not include any
dressing.
{\bf (b)} Same as Fig.3(a) but $|n|=2$.
{\bf (c)} Same as Fig.3(a) but $|n|=3$.
{\bf (d)} Same as Fig.3(a) but $|n|=4$.}
\end{figure}

We note that in the forward direction the differential cross sections
are comparable for first order processes, $n=\pm1$ and $\pm2$;
for second order processes,
$n=\pm3$ and $\pm4$, they are at least three orders of magnitude smaller.

Also displayed in these figures are the results based on the
first order dressing of the target. This approximation
implies that the transition matrix element in Eq.(\ref{gen}) contains
only the first two terms. By inspecting Figs.3 one can see that this
approximation (dashed line) is very good for $|n|\leq2$,
but it can not reproduce the
minimum structure of the differential cross sections for $n=3,4$.

We note also that the closure approximation (dotted-dashed line)
is excellent for $n=\pm1$ and quite fair for $n=\pm2$.
As expected, the electronic contribution alone (dotted line)
fails at small scattering angles.

We have also investigated the situation in which the two components of
the bichromatic field have different intensities. We have chosen to illustrate
the intensity dependence of the angular distribution for the second pair of
sidebands, $n=\pm2$. In Fig.4 three different values of the harmonic intensity
are considered but the intensity of the fundamental is the same,
$I_1=10^{13}$W/cm$^2$.
The harmonic intensities correspond to three values
of the ratio $\rm f =I_2/I_1$, namely $\rm f $=1 represented by full lines,
$\rm f$ =0.1 by dotted-dashed lines, and $\rm f$ =0.01 by dotted lines.

In Fig.4(a) the field components are in phase, $\varphi=0$,
and in Fig.4(b) they are out of phase, $\varphi=\pi$.
The interferences effects are significant for small
values of $\rm f$, as can be understood from
Eq.(\ref{sec2}), because in this case the second order processes are due to fields
which are much stronger than the harmonic field that gives first order processes.
Comparing the graphs in Figs.4(a) and (b) one can see the effect of the relative
phase, too.
\begin{figure}
\includegraphics[width=6.5in,angle=0]{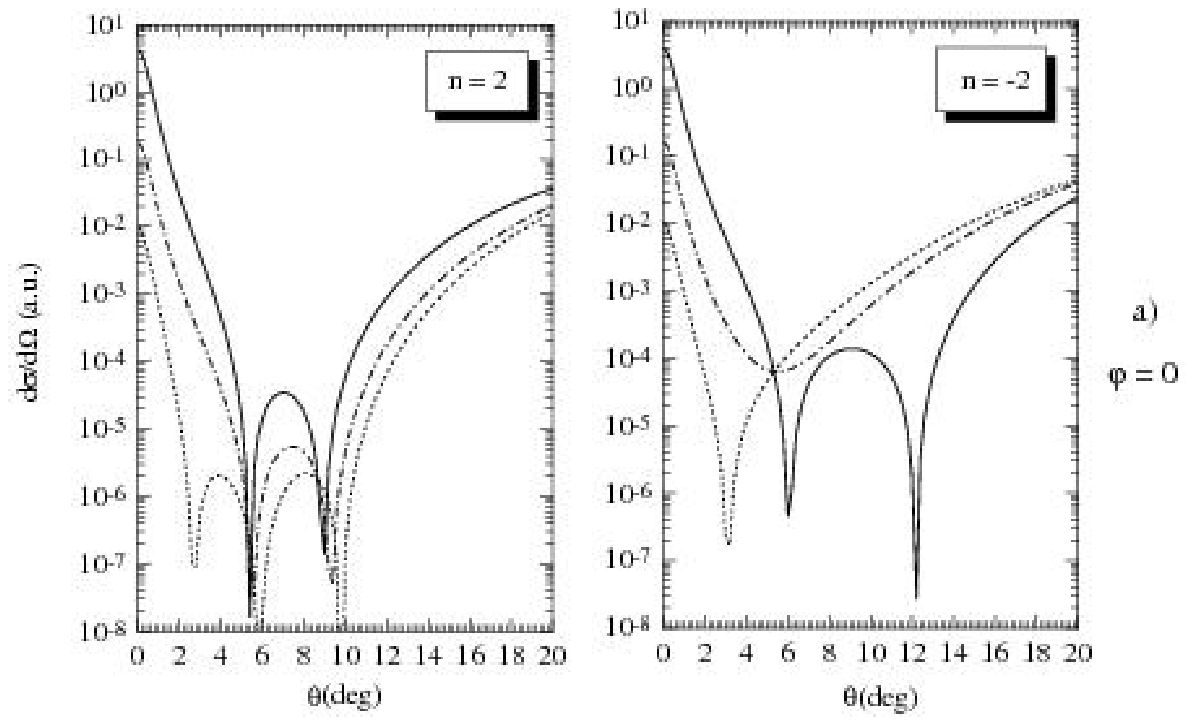}\\
\includegraphics[width=4.5in,angle=0]{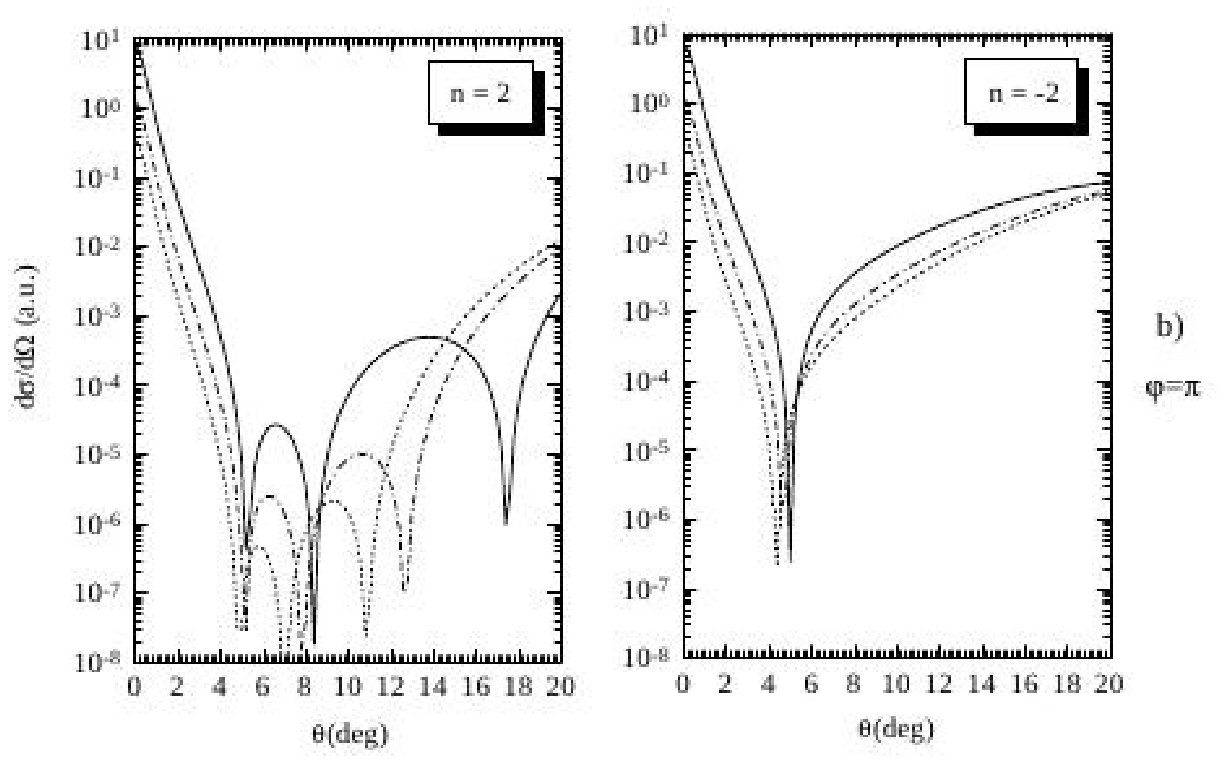}
\caption{
{\bf Fig.4(a)} Differential cross sections, based on Eq.(\ref{gen}), for $|n|=2$
as a function of the scattering angle $\theta $ at the initial energy
$E_i=100$ eV for Nd:YAG laser. The intensity of the fundamental is $I_1=10^{13}$ W/cm$^2$
and the fields are in phase; the harmonic intensity corresponds to the following
cases: $\rm f=1$ (full line), $\rm f=0.1$ (dotted-dashed line), and
$\rm f=0.01$ (dotted line).
{\bf Fig.4(b)} Same as Fig.4(a) but $\varphi=\pi$.}
\end{figure}

Figs.5 display the $\varphi $-dependence of the
laser assisted signal for $n=2$ at two scattering angles: $\theta = 2^0$
in Fig.5(a) and $\theta = 7^0$ in Fig.5(b). The fundamental and the
harmonic have the same intensities as in Fig.4.
At $\theta =2^0$ the laser assisted
signals have their minimal value when the fields are in phase.
The situation is different for $\theta =7^0$: the signals decrease when the
fields are out of phase.
When the perturbative treatment is valid,
for a fixed scattering angle, the difference between the laser assisted signals
at $\varphi=0$ and $\varphi=\pi $ is proportional to
$4 \sqrt{{\rm f} I_1^3}{\cal T}_c{\cal T}_d $ and decreases
for weaker harmonic intensities.
Nevertheless, as discussed earlier, the interference between first and second order
processes is stronger in this case.
The differential cross sections in Eqs.(\ref{sec1}, \ref{sec2}) have different
formal dependences on the intensities of the two components of the bichromatic
field. This explains why the interferences
between first and second order processes and the phase effects
are both stronger for stronger harmonics when $n=\pm 1$. Note however that
the interference effects are significantly affected whenever the frequencies match an
atomic resonance.

\begin{figure}
\includegraphics[width=6in,angle=0]{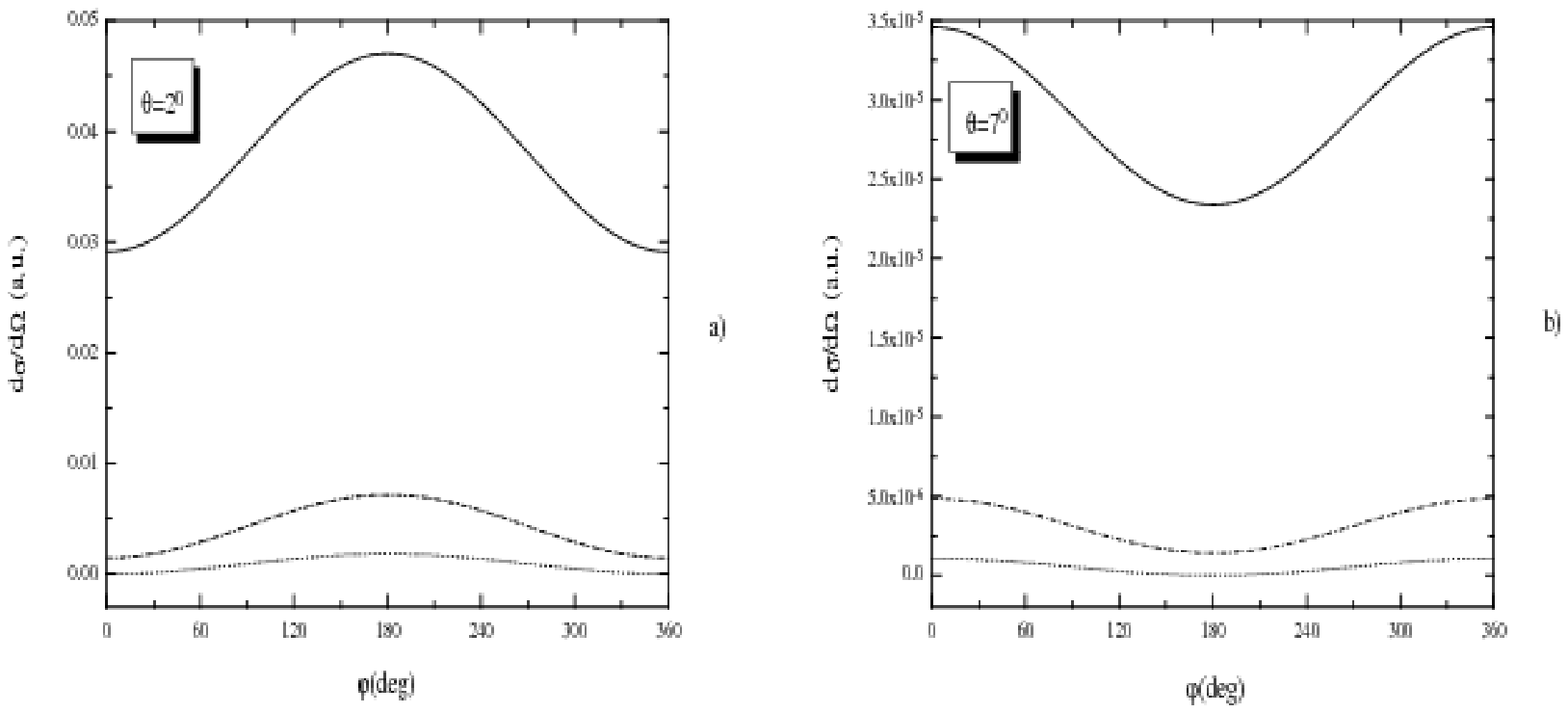}
\caption{
{\bf Fig.5(a)} Differential cross sections, based on Eq.(\ref{gen}), for $n=2$
as a function of the relative phase $\varphi $ at the scattering energy
$E_i=100$ eV for Nd:YAG laser. The scattering angle is $\theta=2^0$ and the fields
intensities are the same as in Fig.4.
{\bf Fig.5(b)} Same as Fig.5(a) but $\theta=7^0$.}
\end{figure}

\subsection{$\omega$ = 4 eV}

\begin{figure}
\includegraphics[width=4.5in,angle=0]{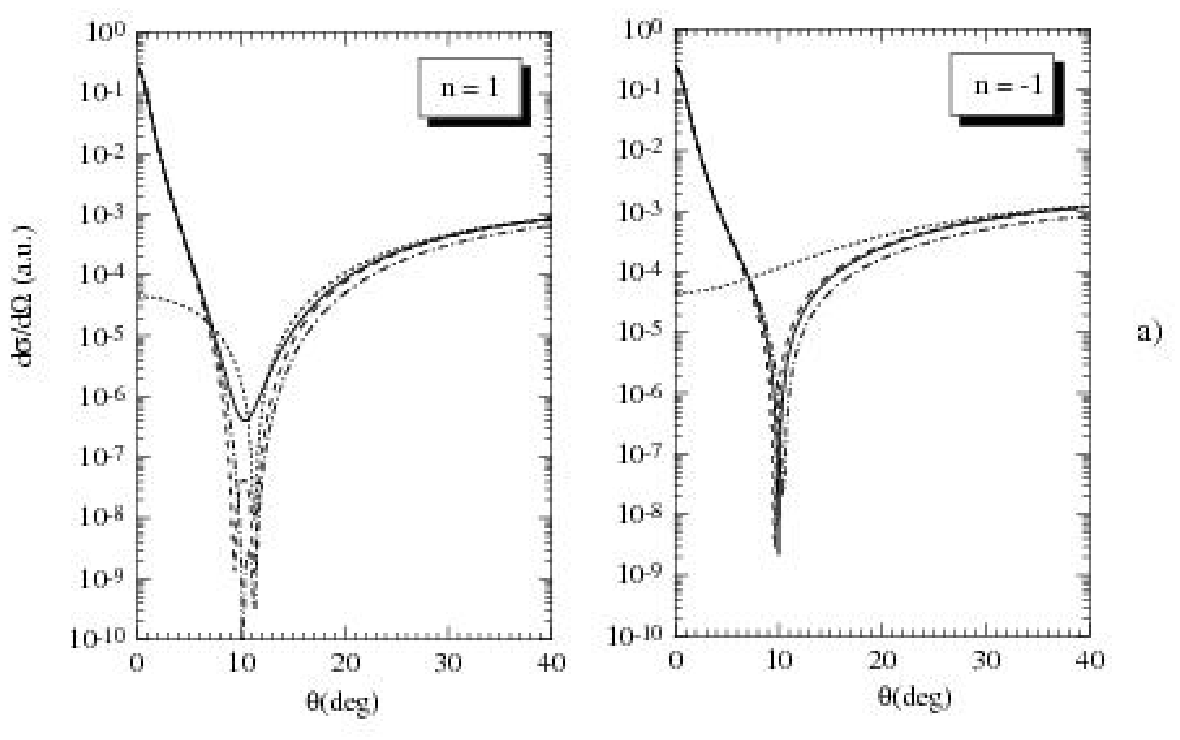}\\
\includegraphics[width=4.2in,angle=0]{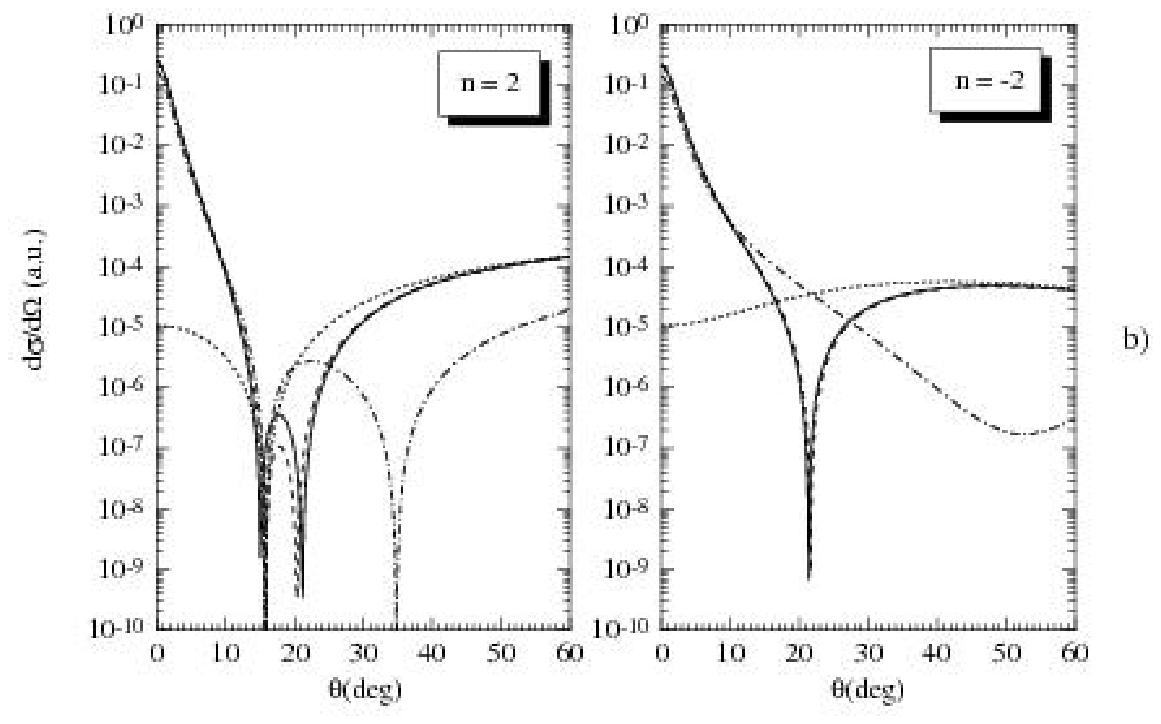}\\
\includegraphics[width=4in,angle=0]{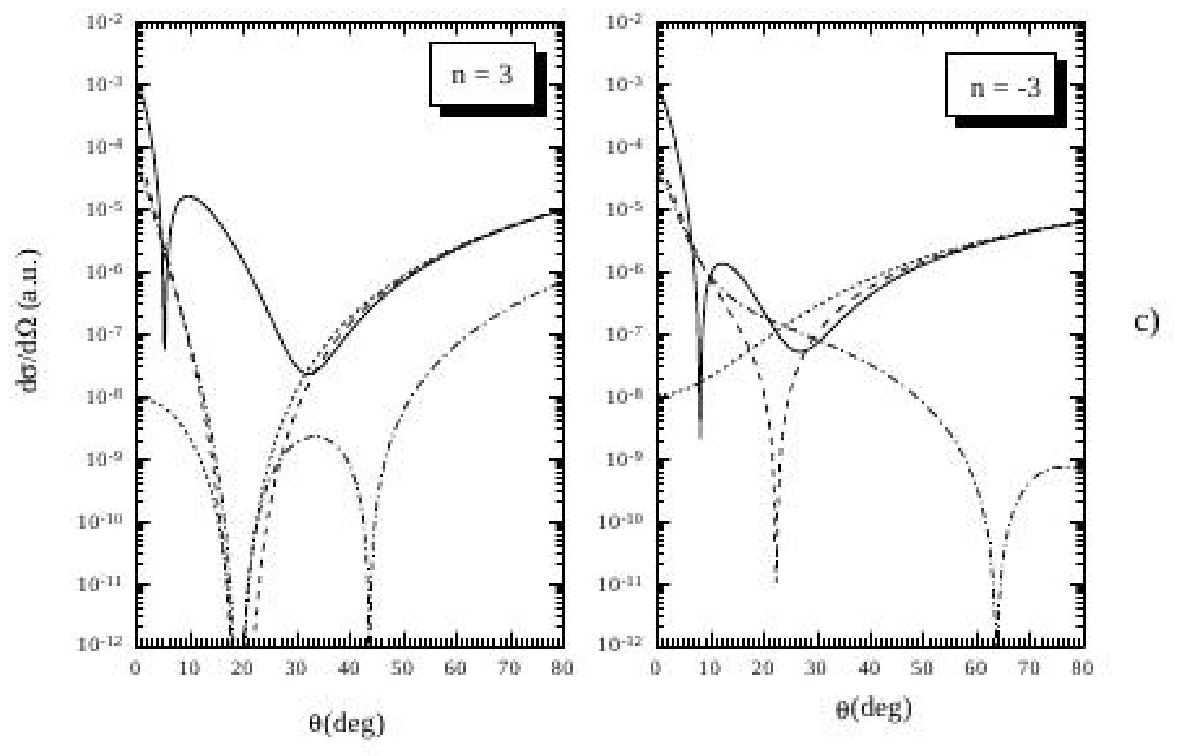}
\caption{
{\bf Fig.6(a-c)} Same as Fig.3(a-c) but $\omega_1=4$ eV.}
\end{figure}

Our interest for higher frequencies is due to the fact that, on one hand,
the dressing of the target is more important than for low frequencies and,
on the other hand, the quiver amplitude is smaller, which increases the
$\theta$-domain for which one can successfully apply the perturbation theory.
Figs.6 show the differential cross sections for the first three pairs of sidebands,
$|n|\leq 3$ at the fundamental frequency
$\omega_1=4$ eV, when the field components are in phase,
$\varphi =$0. The other conditions, concerning the geometry, the scattering
energy, and the intensities, are the same as in Figs.3.
The graphs in Fig.6(a), corresponding to $n=\pm1$, show that
the approximation based on the
first order atomic dressing (dashed line) is quite good in the forward
direction ($\theta < 5^0$) even
in the closure approximation
(dotted-dashed line). On the contrary, for $n=\pm2$, in Fig.6(b), the
closure approximation fails to give fair results for $\theta>15^0$,
but the first order target dressing
leads again to rather good results if the
atomic matrix elements (\ref{defm1}) are evaluated exactly.
The situation is completely different for $n=\pm3$: in Fig.6(c)
the first order dressing gives results which are completely inadequate
for scattering angles smaller than $\theta=30^0$.
In addition, at large scattering angles,
the closure approximation
significantly overestimates the matrix elements (\ref{defm1}), which
affects considerably the differential cross sections.

\section{Conclusions}
Our investigations of free-free transitions in a bichromatic field at
moderate intensities show that calculations based on a perturbative
description of the target in the field are very helpful.
At low frequencies, first order corrections account for the major
dressing effects.
For those situation in which these corrections give reliable
descriptions we investigate the validity of the closure approximation.
We warn that its use should be limited to low
frequencies and small scattering angles.
We stress that, whenever the dominant process
involves two photons,
second order corrections are very important, especially at high frequencies.
These corrections influence the angular distributions and the phase dependences
of the laser assisted signals.

\vspace*{0.25cm}
{\bf Acknowledgments}

This research is supported in part by a Grant of the
Romanian Ministry of Research.

\vspace*{0.25cm}
{\bf Figure Captions}
\vspace*{0.25cm}

{\bf Fig.1(a)} Feynman diagrams corresponding to
one photon absorption by the projectile. The single line represents the
free- and the double one the bound-electron. The horizontal line
denotes the projectile-target interaction.
{\bf Fig.1(b)} Feynman diagrams describing the absorption of $\omega_2$
and the emission of $\omega_1$ by the free electron.
{\bf Fig.1(c)} Same as Fig.1(a) but for the bound electron.
{\bf Fig.1(d)} Feynman diagrams in which each electron
interacts with one photon.
{\bf Fig.1(e)} Same as Fig.1(b) but for the bound electron.

{\bf Fig.2(a-b)} Channels leading to the final energy
$E_f=E_i+n\omega_1$ with $n=1$.
{\bf Fig.2(c-d)} Same as Fig.2(a) but $n=2$.
{\bf Fig.2(e)} Same as Fig.2(a) but $n=3$.
{\bf Fig.2(f)} Same as Fig.2(a) but $n=4$.

{\bf Fig.3(a)} Differential cross sections for $|n|=1$ as a function of the scattering
angle $\theta $ at the initial energy
$E_i=100$ eV for Nd:YAG laser ($\omega_1$=1.17 eV). $I_1= I_2=10^{12}$ W/cm$^2$
and the fields are in phase, $\varphi=0$. Full lines represent the results based on
Eq.(\ref{gen}), dotted lines include only first order dressing, dotted-dashed lines
correspond to the closure approximation, dotted lines do not include any dressing.
{\bf Fig.3(b)} Same as Fig.3(a) but $|n|=2$.
{\bf Fig.3(c)} Same as Fig.3(a) but $|n|=3$.
{\bf Fig.3(d)} Same as Fig.3(a) but $|n|=4$.

{\bf Fig.4(a)} Differential cross sections, based on Eq.(\ref{gen}), for $|n|=2$
as a function of the scattering angle $\theta $ at the initial energy
$E_i=100$ eV for Nd:YAG laser. The intensity of the fundamental is $I_1=10^{13}$ W/cm$^2$
and the fields are in phase; the harmonic intensity corresponds to the following
cases: $\rm f=1$ (full line), $\rm f=0.1$ (dotted-dashed line), and
$\rm f=0.01$ (dotted line).
{\bf Fig.4(b)} Same as Fig.4(a) but $\varphi=\pi$.

{\bf Fig.5(a)} Differential cross sections, based on Eq.(\ref{gen}), for $n=2$
as a function of the relative phase $\varphi $ at the scattering energy
$E_i=100$ eV for Nd:YAG laser. The scattering angle is $\theta=2^0$ and the fields
intensities are the same as in Fig.4.
{\bf Fig.5(b)} Same as Fig.5(a) but $\theta=7^0$.

{\bf Fig.6(a-c)} Same as Fig.3(a-c) but $\omega_1=4$ eV.
\end{document}